%% file: RhOmegaMixInJpsiVP_wangdong_arxiv_v1.tex
\documentclass[%
 reprint,
 amsmath,amssymb,
 aps,
]{revtex4-1}

\usepackage{graphicx}% Include figure files
\usepackage{dcolumn}% Align table columns on decimal point
\usepackage{bm}% bold math
\usepackage[colorlinks,linkcolor=blue,anchorcolor=blue,citecolor=green,urlcolor=magenta,
	linktocpage=true,breaklinks=true,nesting=true,pdfpagemode=UseThumbs]{hyperref}
\usepackage{breakurl} %it is should be used after usepackage"hyperref"
\usepackage{ulem}  %used to add underline Note: this will conflict with "\cite{}"
\usepackage[tight]{subfigure}
\usepackage{pdfpages}
\usepackage{colortbl}
\usepackage{supertabular}

\def \jpsi {J/\psi}

\def \rhopipipi   {\rho^{0}\to\pi^{+}\pi^{-}\pi^{0}}
\def \omegapipipi {\omega\to\pi^{+}\pi^{-}\pi^{0}}
\def \jpsirhopi   {J/\psi\to\rho^{0}\pi^{0}}
\def \jpsiomegapi {J/\psi\to\omega\pi^{0}}
\def \jpsiomegarhopipipipi {J/\psi\to\omega(\rho)\pi^{0}\to\pi^{+}\pi^{-}\pi^{0}\pi^{0}}

\def \jpsiVpipipipi {J/\psi\to V\pi^{0}\to\pi^{+}\pi^{-}\pi^{0}\pi^{0}}

\allowdisplaybreaks[3] %to make the formulas page

\begin{document}
%\begin{CJK*}{GBK}{song}
\input{./charpters/TittleAbstract20130611}

%\begin{multicols}{2}

\input{./charpters/Introduction20130605}

\input{./charpters/Jpsi4pi20130605}
\input{./charpters/FitPDG20130605}

\input{./charpters/Conclusion20130605}

\acknowledgments{The authors are grateful to Yuan Changzheng for the initiation of this paper, and to Zheng Hanqing, Zhao Qiang, Shi Meng,  Rinaldo Baldini Ferroli and Rafel Escribano for the helpful discussions and suggestions during the research. The work is supported by the Ministry of Science and Technology of China under ``973 Project No. 2009CB825200".}

%\end{multicols}

\vspace{10mm}
%\newpage
%\begin{multicols}{2}
\input{./charpters/Appendixes20130605}

%\end{multicols}

\vspace{-1mm}
\centerline{\rule{80mm}{0.1pt}}
\vspace{2mm}

%\begin{multicols}{2}
%%%%%%%References
%\input{./charpters/References20130415}
\bibliographystyle{apsrev4-1}
\bibliography{./ref/library20130605}
%\end{multicols}

%\newpage
%\input{charpters/VersionInfo20130605}

\clearpage
%\end{CJK*}
\end{document}

%% file: charpters/TittleAbstract20130611.tex
%\fancyhead[c]{\small Chinese Physics C~~~Vol. 37, No. 1 (2013)
%010201} \fancyfoot[C]{\small 010201-\thepage}
%
%\footnotetext[0]{Received 14 March 2009}

\title{$\rho - \omega$ Mixing in $\jpsi\to VP$ Decays \thanks{Supported by Ministry of Science and Technology of China (973 Project No. 2009CB825200)}}

\author{%
      WANG Dong$^{1}$%
\quad BAN Yong$^{1;1)}$\email{bany@pku.edu.cn}%
\quad LI Gang$^{2}$
}

\address{%
$^1$ School of Physics, State Key Laboratory of Nuclear Physics and Technology, Peking University,  Beijing 100871, China\\
$^2$ Institute of High Energy Physics, Chinese Academy of Sciences, Beijing 100049, China\\
}

\begin{abstract}
The study on $\rho-\omega$ mixing is mainly focused on vector mesons decay with isospin $I=1$, namely $\rho(\omega)\to\pi^+\pi^-$ process. In this paper, we present the study of $\rho-\omega$ mixing in $\rho(\omega)\to\pi^+\pi^-\pi^0$ ($I=0$) using a flavor parameterization model for $J/\psi\to VP$ process. By fitting theoretical frame to PDG data, we obtain the SU(3)-breaking effect parameters $s_V=0.03\pm 0.12,\ s_P=0.17\pm 0.17$ and the $\rho-\omega$ mixing polarization operator $\Pi _{\rho \omega }=0.006\pm 0.011\text{ \text{GeV}}^2$. The branching ratios are also renewed when mixing effect is incorporated:
$Br(J/\psi\to \omega\pi^0) = (3.64 \pm 0.37)\times 10^{-4}$,
$Br(J/\psi\to \omega\eta)  = (1.48 \pm 0.17)\times 10^{-3}$,
$Br(J/\psi\to \omega\eta^{\prime}) = (1.55\pm 0.56)\times 10^{-4}$;
 they are different from the corresponding PDG2012 values by $19\%$, $15\%$ and $15\%$, respectively.
\end{abstract}

%\begin{keyword}
%$\rho-\omega$ mixing, branching ratio, $J/\psi\to VP$ decay.
%\end{keyword}

\begin{pacs}
12.39.-x,
12.40.Vv,
13.25.Gv.\\
Submitted to Chinese Physics C (CPC).
\end{pacs}

\maketitle

%\footnotetext[0]{\hspace*{-3mm}\raisebox{0.3ex}{$\scriptstyle\copyright$}2013
%Chinese Physical Society and the Institute of High Energy Physics
%of the Chinese Academy of Sciences and the Institute
%of Modern Physics of the Chinese Academy of Sciences and IOP Publishing Ltd}%

%% file: charpters/Introduction20130605.tex
\section{INTRODUCTION}
In 1961, Glashow suggested that electromagnetic transition leads to $\rho-\omega$ mixing~\cite{PhysRevLett.7.469}. Eight years later, a direct experimental evidence for $\rho-\omega$ mixing was observed~\cite{Augustin:1969hk}, in the next year Willemsen followed up the study~\cite{PhysRevD.2.133}. In the following thirty years, along with the development of VMD (vector meson dominance) model~\cite{PhysRevLett.7.426,PhysRevLett.22.981,OConnell1997a,OConnell1997,Feynman1998}, many theories were proposed to understand $\rho-\omega$ mixing, such as CVS (Charge Symmetry Violation)~\cite{McNamee1975483,arxiv9507010,Goldman1992,PhysRevC.52.3428}, QCDSR (Quantum Chromodynamics Sum Rules)~\cite{Shifman1979519,PhysRevD.53.2563}, ChPT (Chiral Perturbation Theory)~\cite{Urech1995308,arxiv9502393} and HLS (Hidden Local Symmetry)~\cite{Benayoun2009,Benayoun2008,Benayoun2001}.

Up to present, most of $\rho-\omega$ mixing studies are based on vector meson decays with isospin $I=1$, namely, $\rho(\omega)\to\pi\pi$.
The mixing in isospin $I=1$ transition has been well studied both theoretically and experimentally~\cite{Bramon1986, OConnell1997, SGandHBOC1998}.

However, the mixing in $\rho(\omega)\to3\pi$ decay with isospin $I=0$ is not so well understood yet. Because $\Gamma_{\rho}\gg(m_{\omega}-m_{\rho})$ and $Br(\omegapipipi)\gg Br(\rhopipipi)$, it is difficult to measure the process directly from experiment~\cite{LiuFang2004}. A study on $\rho(\omega)\to3\pi$ interference with $\jpsi\to VP$ decay has been made using a flavor parameterization method~\cite{Bramon1986,LiuFang2004}. With $J/\psi$ decays, the small value of $Br(\rhopipipi)/Br(\omegapipipi)$ can be compensated by the large value of $Br(\jpsirhopi)/Br(\jpsiomegapi)$ in some extent, which provides a new insight in $\rho-\omega$ mixing study. The parameterization of $\jpsi\to VP$ process has been developed with single and double Okubo-Zweig-Iziuka (SOZI, DOZI) rules~\cite{Abraham1988,Escribano2010,Kopke1989,Li2008,Thomas2007}.

SND group has taken $\rho-\omega$ mixing effect into account in the study of $e^+e^-\to 3\pi$ decay below $0.98 \text{GeV}$~\cite{Achasov2003}. Its theoretical model with $e^+e^-\to 3\pi$ may also be considered in the study of $e^+e^-\to\jpsiomegarhopipipipi$.

Mixing phenomenon between $\rho$ and $\omega$ in $\jpsi$ decays will serve as an important probe for the test of various theoretical models and the G-parity violation. The main purpose of this paper is trying to study $\rho-\omega$ mixing with a flavor parameterization method in $J/\psi\to VP$ decay. We expect to derive the mixing parameter $\prod_{\rho\omega}$, and to modulate the measured $Br(J/\psi\to\omega\pi^0)$ and $Br(J/\psi\to\omega\eta(\eta^{\prime}))$ according to the mixing value.

The contents of our paper are organized as follows. In section 2, referring to $e^+e^-\to 3\pi$ process~\cite{Achasov2003} and taking into account of the $\omegapipipi$ contact term~\cite{DGarciaGudino2012}, we describe the process of $e^{+}e^{-}\to \jpsiVpipipipi$ and give its cross section. In section 3, by using the flavor parameterization method~\cite{Abraham1988,Escribano2010}, we perform a fit with the theoretical frame of $\jpsi\to VP$ decay to the existing data. The conclusion and the interpretation of the results are given in section 4. The appendixes are devoted to the detailed notations in mixing formulae.

%% file: charpters/Jpsi4pi20130605.tex
\section{THEORETICAL FRAME OF $e^{+}e^{-}\to \jpsiVpipipipi$ PROCESS}\label{sec:jpsi4pi}

The SND result~\cite{Achasov2003} and related branching ratios in PDG2012~\cite{PDG2012} indicate that the decay channels $\jpsi\to\rho^{\prime}\pi^0,\jpsi\to\rho^{\prime\prime}\pi^0,\jpsi\to\omega^{\prime}\pi^0,\jpsi\to\omega^{\prime\prime}\pi^0$
have little contribution in our interested process. We will omit these channels and calculate $e^{+}e^{-}\to \jpsiVpipipipi (V=\rho,\omega,\phi)$ process in this paper.
The framework used by SND~\cite{Achasov2002,Achasov2003,Achasov2002a,Achasov2005} is adopted in the calculation, and the $\omegapipipi$ contact term is taken into account~\cite{DGarciaGudino2012,Aviv1972,Kuraev1995,LucioM2000,OKaymakcalanSRajeev1984,Rudaz1984}.

The cross section of $e^{+}e^{-}\to \jpsiVpipipipi$ process is
\begin{equation}\label{eq:dsigmadm}
\begin{split}
&\frac{d\sigma \left(s,m_0,m_+\right)}{dm_0dm_+}=\frac{1}{s^{3/2}}\frac{\left|\overset{\rightharpoonup }{p_+}\times \overset{\rightharpoonup }{p_-}\right|^2}{12\pi ^2\sqrt{s}}m_0m_+|F|^2  , \\
&F=F_{\rho \pi }(s)+F_{\omega \pi }(s)+F_{3\pi }(s) .
\end{split}
\end{equation}
Here $s$ is the invariant mass of $\pi^{+}\pi^{-}\pi^{0}$ system, $\overset{\rightharpoonup }{p_+}$ and $\overset{\rightharpoonup }{p_-}$ are the momenta of $\pi^+$ and $\pi^-$ mesons in the $3\pi$ system rest frame. $m_+$ and $m_0$ are the invariant masses of $\pi^{+}\pi^{0}$ and $\pi^{+}\pi^{-}$.

\begin{figure*}[!htbp]
\centering
\subfigure[\small{$\rho\pi$ channel contributions.}]
{\includegraphics[scale=1.0,height=0.1\textheight,width=0.30\textwidth]
{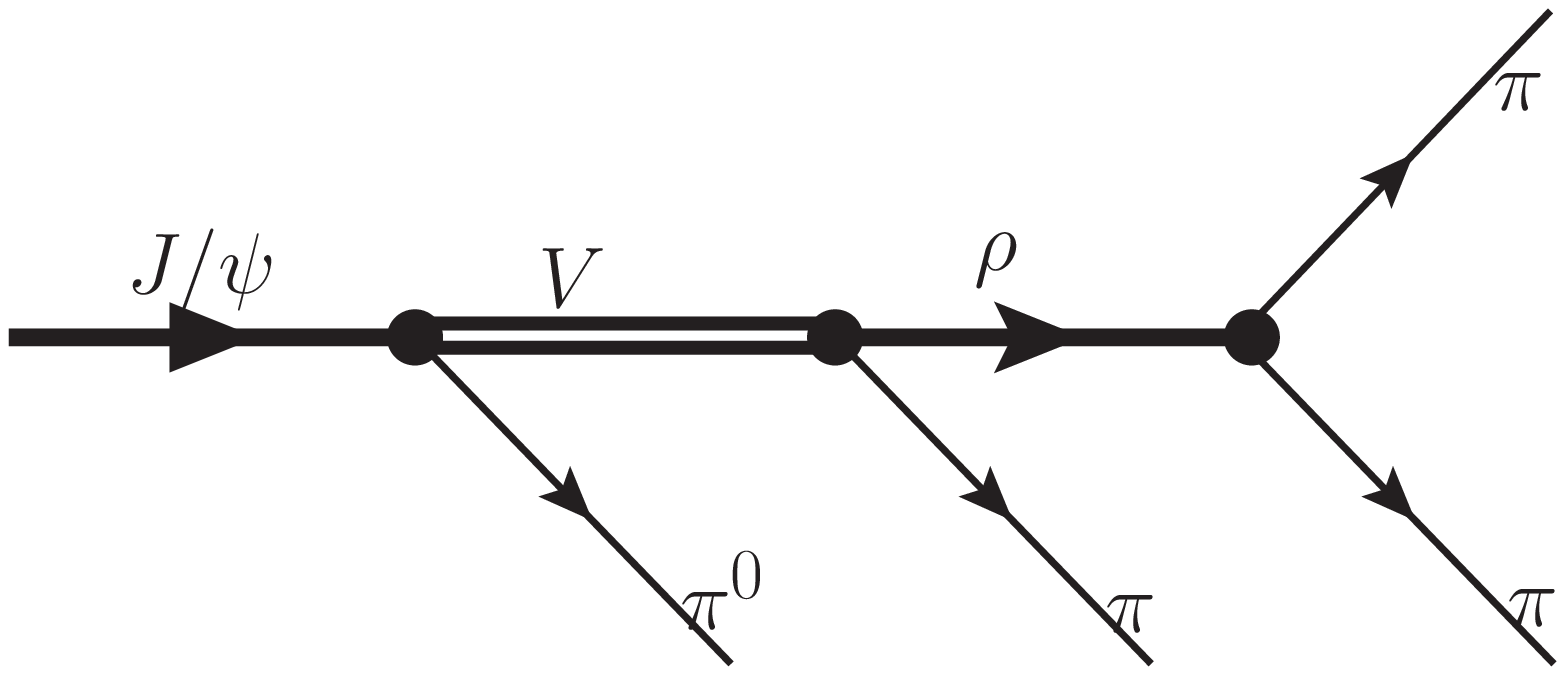}}
\quad
\subfigure[\small{possible transition $V\to\rho^{\prime(\prime\prime)}\pi\to\pi^{+}\pi^{-}\pi^{0}$.}]
{\includegraphics[scale=1.0,height=0.1\textheight,width=0.30\textwidth]
{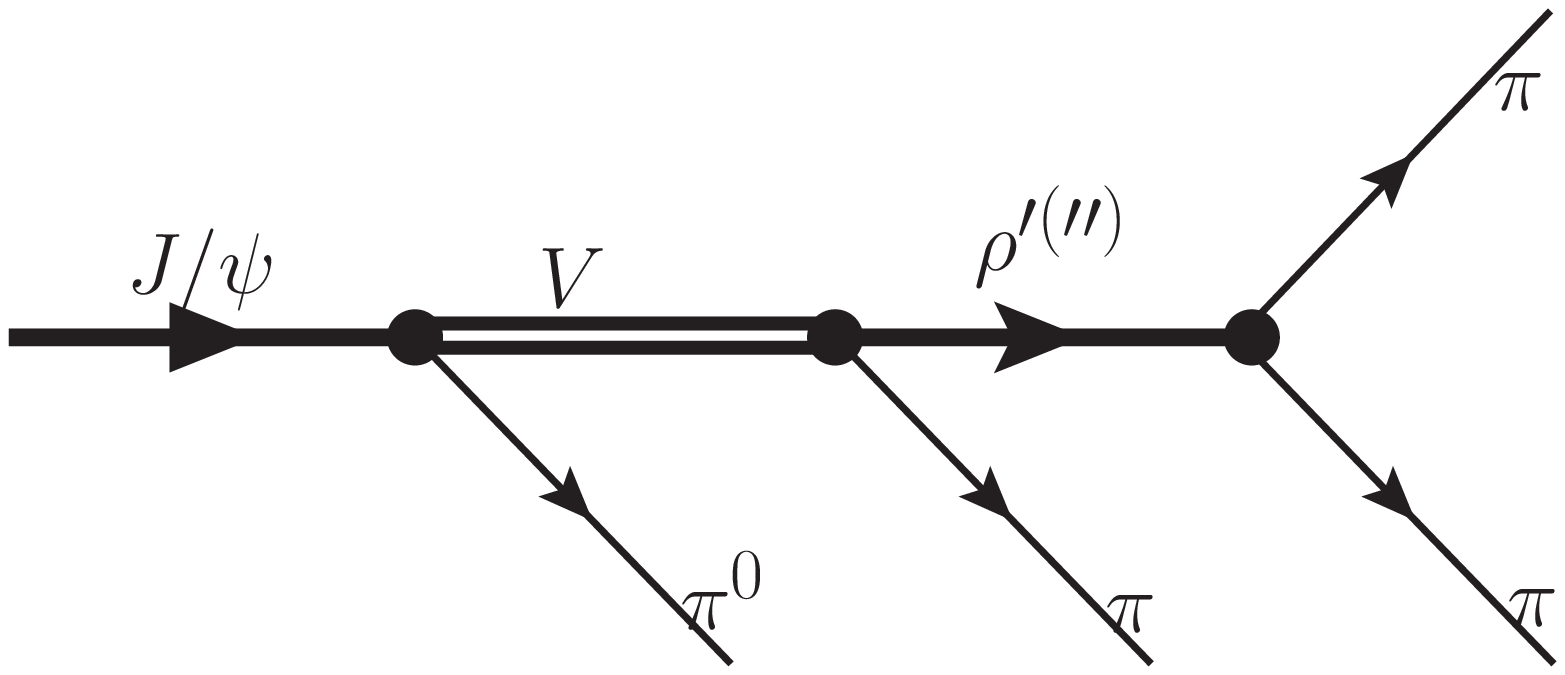}}
\quad
\subfigure[\small{interaction of $\rho$ and $\pi$ mesons in the final state.}]
{\includegraphics[scale=1.0,height=0.1\textheight,width=0.30\textwidth]
{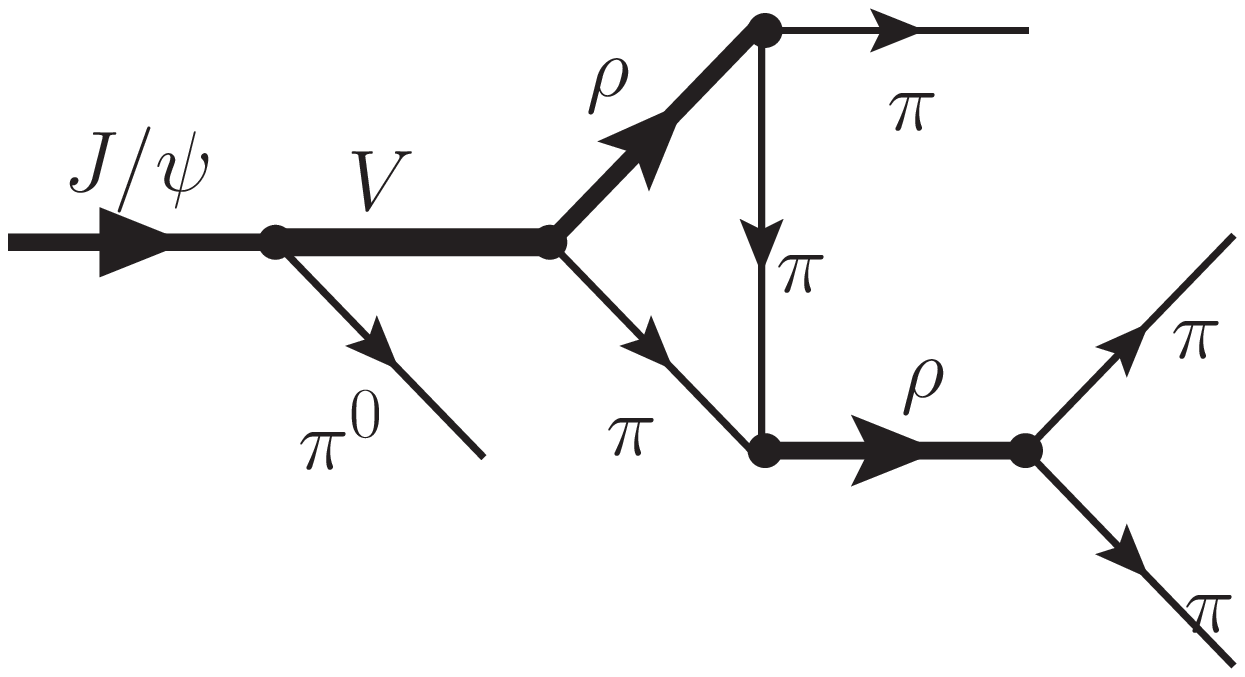}}\\
\subfigure[\small{$\omega\pi$ channel contributions.}]
{\includegraphics[scale=1.0,height=0.1\textheight,width=0.30\textwidth]
{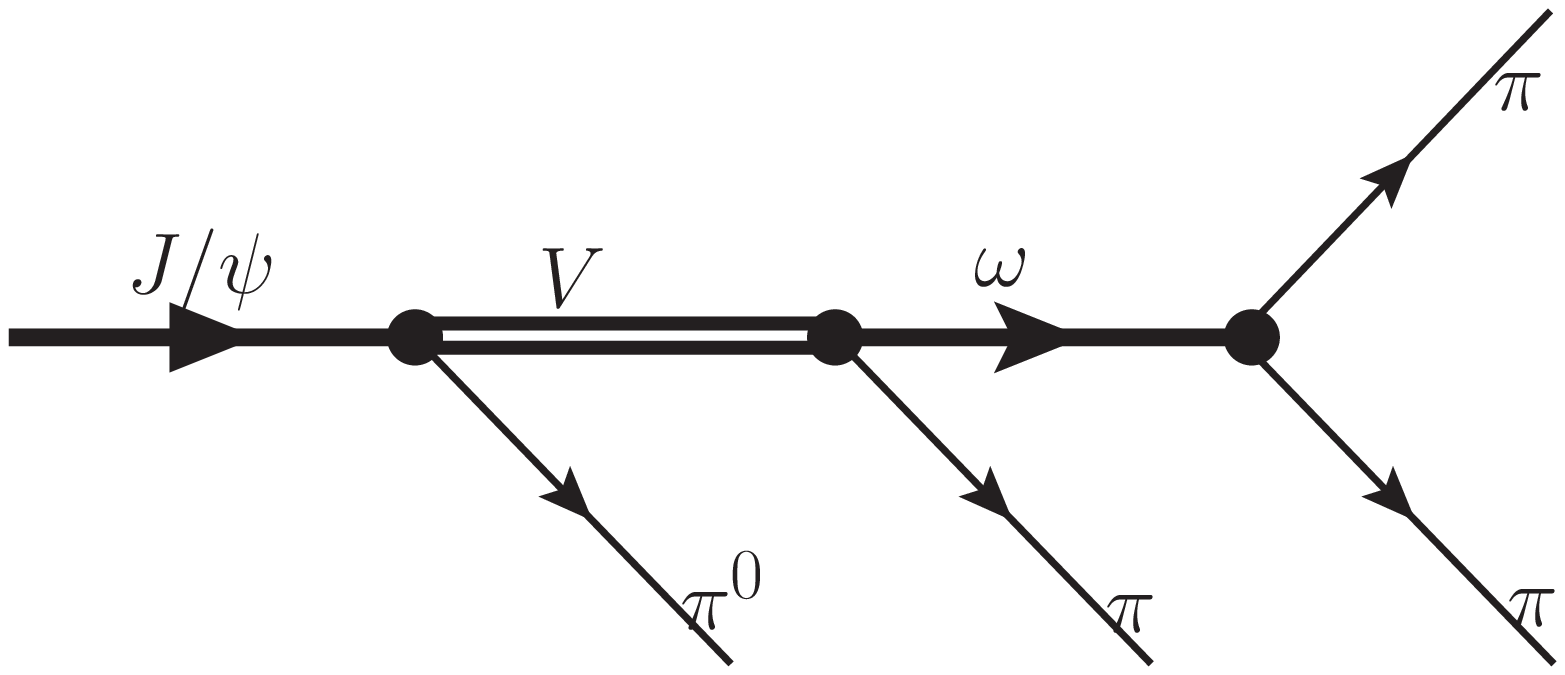}}
\quad
\subfigure[\small{contact term for higher order contributions, which requires the same space-time point for all particles when decay happens.}]
{\includegraphics[scale=1.0,height=0.1\textheight,width=0.30\textwidth]
{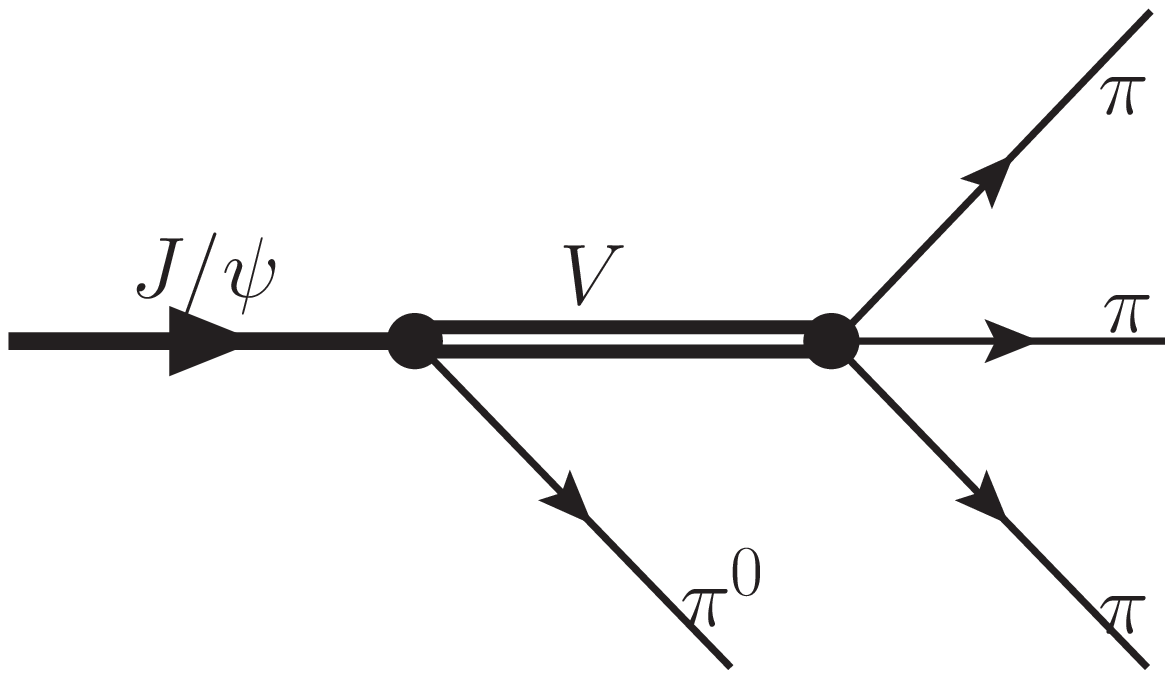}}
\quad\hspace{0.30\textwidth}
\caption{\label{fig:decays}$\jpsiVpipipipi$ process.}
\end{figure*}

$F_{\rho \pi }(s)\ (F_{\omega \pi }(s), F_{3\pi }(s))$ in Eq.~(\ref{eq:dsigmadm}) is the form factor for the vector mesons decays through $V\to \rho \pi\ (V\to \omega \pi, V\to 3\pi)$ channel, taking into account the transition described in Fig.~\ref{fig:decays}(a,b,c) (Fig.~\ref{fig:decays}(d), Fig.~\ref{fig:decays}(e)). They have the forms
\begin{equation} \label{eq:my3f}
\begin{split}
&F_{\rho \pi }(s)=\left[a_{3\pi }+\sum _{i=+,0,-} \frac{g_{\rho ^i\pi \pi }}{D_{\rho ^i}\left(m_i\right)Z\left(m_i\right)}\right] \times\\
&\qquad \qquad \{2g_{\omega \rho \pi }(s)\left[\frac{A_{\psi \omega \pi }(s)}{D_{\omega }(s)}-\frac{\Pi _{\rho \omega }A_{\psi \rho \pi }(s)}{D_{\omega }(s)D_{\rho }(s)}\right]+ \\
&\qquad \qquad \frac{2A_{\psi \phi\pi }(s)g_{\phi\rho \pi }e^{\text{i}\phi_{\omega \phi}}}{D_{\phi}(s)}\},\\
&F_{\omega \pi }(s)=\frac{-\Pi _{\rho \omega }g_{\rho ^0\pi \pi }}{D_{\omega }(m_0)D_{\rho }(m_0)}\frac{2A_{\psi \rho\pi}(s)g_{\rho\omega \pi}}{D_{\rho}(s)}, \\
&F_{3\pi }(s)=6g_{\text{$\omega $3$\pi $}}\left[\frac{A_{\psi \omega \pi }(s)}{D_{\omega }(s)}-\frac{\Pi _{\rho \omega }A_{\psi \rho \pi }(s)}{D_{\omega }(s)D_{\rho }(s)}\right].
\end{split}
\end{equation}
%\end{multicols}
%\ruleup
%\ruledown
%\begin{multicols}{2} 

$m_-$ is the invariant mass of $\pi^{-}\pi^{0}$ and satisfy
\begin{equation}
m_-=\sqrt{s+m_{\pi ^0}^2+2m_{\pi }^2-m_0^2-m_+^2}.
\end{equation}

$A_{\psi V \pi}(s)$ is the amplitude for $e^{+}e^{-}\to \jpsi \to V \pi^0$ process
\begin{small}
\begin{equation}\label{eq:ajpsivp2}
\begin{split}
A_{\psi V\pi}(s)\equiv &g\sqrt{Br\left(J/\psi \rightarrow \text{V$\pi $}^0\right)} \times \\
&\sqrt{\frac{q^3\left(m_{\psi },\sqrt{s},m_{\pi ^0}\right)e^{-q^2\left.\left(m_{\psi },\sqrt{s},m_{\pi ^0}\right)\right/8\beta ^2}}{q^3\left(m_{\psi },m_V,m_{\pi ^0}\right)e^{-q^2\left.\left(m_{\psi },m_V,m_{\pi ^0}\right)\right/8\beta ^2}}},
\end{split}
\end{equation}
\end{small}
where $g$ is a factor with a dimension  $\text{GeV}^2$, which includes the coupling constant of decay $e^{+}e^{-}\to \jpsi$; $q$ is the momentum defined as
\begin{footnotesize}
\begin{equation}
\begin{split}
q\left(M,m_1,m_2\right)=\frac{\sqrt{\left[M^2-\left(m_1+m_2\right){}^2\right]\left[M^2-\left(m_1-m_2\right){}^2\right]}}{2M}.
\end{split}
\end{equation}
\end{footnotesize}

$g_{\text{V$\rho(\omega) \pi $}}$ is the coupling constant for decay $V\to\rho(\omega)\pi$. $g_{\omega\pi\pi}$ and $g_{\rho\pi\pi}$ are the coupling constants for decays $\omega\to\pi\pi$ and $\rho\to\pi\pi$, respectively. $g_{\omega 3\pi}$ is the coupling constant for contact term $\omega\to 3\pi$. The values of those coupling constants were calculated according to Refs.~\cite{Achasov2003,Achasov2002,DGarciaGudino2012}: $g_{\rho^0\pi \pi}=5.975$, $g_{\rho^{\pm}\pi \pi}=5.989$, $g_{\rho\omega\pi}=16.8 \text{GeV}^{-1}$, $g_{\omega\rho\pi}=15.0 \text{GeV}^{-1}$, $g_{\phi\rho\pi}=0.827 \text{GeV}^{-1}$, and $g_{\omega 3\pi}=-47.0 \text{GeV}^{-3}$.

$\phi_{\rho V} (\phi_{\omega V})$ is a relative interference phase between $\rho (\omega)$ and vector mesons $V$; thus $\phi_{\rho \rho}=0,\ \phi_{\omega \omega}=0$. We adopt the value $\phi _{\omega \phi }=(163\pm 3 \pm 6)^{\circ}$ obtained by SND~\cite{Achasov2003}, which takes into account of the $\phi-\omega$ mixing and consists with the theoretical prediction~\cite{Achasov2000}.
%The interference with $\phi\pi^{0}$ is regarded in CMD-2~\cite{RRAkhmetshinetalCMD2004,RRAkhmetshinetalCMD2000}, but without consideration of $\rho^{0}\pi^{0}$ and other channels.

$D_{V}(s)$ is the propagator function defined as
\begin{equation} \label{eq:dv}
D_V(s)=m_V^2-s-i\sqrt{s}\Gamma _V(s),
\end{equation}
where the $s$-dependent widths of vector mesons $\Gamma _V(s)$ are defined in SND~\cite{Achasov2003}.

$a_{3\pi }=(0.1\pm 2.3 \pm 2.5)\ \text{GeV}^{-2}$ represent the contribution from $V\to\rho^{\prime}(\rho~{\prime\prime})\pi\to\pi^{+}\pi^{-}\pi^{0}$ processes~\cite{Achasov2003}. The factor $Z\left(m_i,s\right)$ is defined as~\cite{Achasov1994} $Z(m,s)=1-i s_1\Phi (m,s)$, where $s_1=0.3\pm 0.3\pm 0.3$~\cite{Achasov2003}.

Here the  $\rho-\omega$ mixing in $\jpsi\to (\rho^{0}, \omega)\pi^{0} \to \rho\pi\pi\ (\omega\pi\pi)$ and $V\to (\rho^{0}, \omega) \pi^{0} \to 3\pi$ decays is considered (refer to Eqs.~(\ref{eq:mixing1})(\ref{eq:mixing2})(\ref{eq:mixing3})(\ref{eq:mixing4}) in Appendix A).
It is a well-known fact that the real part of the coupling constant of the direct transition $\omega\to\pi^{+}\pi^{-}$ has no contribution to the amplitude of $\omega\to\pi^{+}\pi^{-}$ decay~\cite{Achasov1994}, therefore we have ignored the term $g_{\omega \pi \pi }^{(0)}$, as well as the terms $g_{\rho \rho \pi }^{(0)}$, or  $g_{\rho 3\pi }^{(0)}$.
$\prod_{\rho\omega}$ is a polarization operator, it is speculated that $\prod_{\rho\omega}$ satisfies: $\text{Im}(\prod_{\rho\omega})\ll \text{Re}(\prod_{\rho\omega})$~\cite{Achasov2003,Achasov1994,Achasov1978}, thus we only consider the real part of $\prod_{\rho\omega}$. Its value should be positive because it is extracted from the module of the amplitude.

The cross section $\sigma(s)$ is defined as the integral of $\sigma \left(s,m_0,m_+\right)$ over $m_0$ and $m_+$:
\begin{equation} \label{eq:sigma0}
\sigma (s)=\int \int \frac{1}{s^{3/2}}\frac{\left|\overset{\rightharpoonup }{p_+}\times \overset{\rightharpoonup }{p_-}\right|^2}{12\pi ^2\sqrt{s}}m_0m_+|F|^2dm_0dm_+ .
\end{equation}

%% file: charpters/FitPDG20130605.tex
\input{./tables/amplitudes1_20130611.tex}

\input{./tables/brvaluesfit_20130611.tex}

%\input{./tables/tabPDGfitresultfinal20130529.tex}
%\begin{multicols}{2}
\section{FIT OF MIXING PARAMETERS}
\subsection{Strategy of the Fit}
A flavor parameterization method used in Ref.~\cite{Abraham1988} is applied here to study the $\jpsi\to VP$ process~\cite{Escribano2010,Li2008,Thomas2007,Kopke1989}. The decays proceed through strong and electromagnetic interaction, where the effects of double Okubo-Zweig-Iziuka (DOZI) rule-violation and SU(3) flavor symmetry breaking should be taking into account.

The general parameterization of the amplitudes is written in Table~\ref{tab:amplitudes1}, where the terms $X_{\eta}$, $Y_{\eta}$, $Z_{\eta}$ and $X_{\eta'}$, $Y_{\eta'}$, $Z_{\eta'}$ include the $\eta - \eta'$ mixing (Eq.~(\ref{eq:etaetapmixing})); $\phi_{P}$ is the $\eta-\eta'$ mixing angle, and $\phi_{\eta'G}$ weights the amount of gluonium in  $\eta'$.
And the terms $\omega_q$, $\phi_s$  mean $\omega-\phi$ mixing is considered as in Eq.~(\ref{eq:omegaphimixing}), then the amplitudes of the decays including $\omega$ or $\phi$ are rewritten as
\begin{equation}
\begin{split}
&M_{\omega}= \cos\theta_{\omega\phi}M_{\omega_q}-\sin\theta_{\omega\phi}M_{\phi_s}, \\
&M_{\phi}  = \sin\theta_{\omega\phi}M_{\omega_q}+\cos\theta_{\omega\phi}M_{\phi_s},
\end{split}
\end{equation}
where $\theta_{\omega\phi}$ is the mixing angle of $\rho$ and $\omega$. The value of $\theta_{\omega\phi}$ can be set to 0 if this mixing is ignored. Here we take the value $\theta_{\omega\phi}=(3.2\pm 0.1)^{\circ}, s_e=0.19\pm 0.05$ as in Refs.~\cite{Escribano2007,Escribano2010,MGandJLR2009}.

Similarly with Refs.~\cite{Zhao2005,Li2008,Thomas2007}, the branching ratio is given by
\begin{equation}\label{eq:brcor}
\begin{split}
Br_{\text{cor}}(J/\psi \to \text{$\omega \pi$})&=\frac{Br(J/\psi \to \omega \pi \to \text{4$\pi $})}{Br(\omega \to 3\pi )} \\
&=\left|M_{\psi \omega \pi^0}\right|^2 q^3 e^{-q^2/8\beta ^2},
\end{split}
\end{equation}
where $\beta$ is a scale of the energy and is commonly adopted to $\beta=0.5 \text{GeV}$~\cite{Zhao2005,Li2008,Thomas2007}.

Discriminated from above ideal branching ratio, the actually measured ratio can be written as
\begin{small}
\begin{equation}\label{eq:bruncor}
Br_{\text{uncor}}(J/\psi \to \text{$\omega \pi$})=\frac{Br(J/\psi \to V \pi \to \text{4$\pi $})}{Br(\omega \to 3\pi )} =f\cdot\sigma_{\pi^0},
\end{equation}
\end{small}
where $\sigma_{\pi^0}$ is the integral of $\sigma(s)$ (Eq.~(\ref{eq:sigma0})), in which $\pi^0$ indicates $\jpsi\to 4\pi$ via $V\pi^0$, instead of via $V\eta$ or $V\eta^{\prime}$. The integrating range is $\sqrt{s}\in [0.6,1.0]$ \text{GeV}. $f$ is a constant factor with dimension  $\text{GeV}^2$ that absorbs the factor of $g$ in Eq.~(\ref{eq:ajpsivp2}), the latter can be redefined as
\begin{small}
\begin{equation}\label{eq:ajpsivp3}
\begin{split}
A_{\psi V\pi}(s)\equiv &\sqrt{Br\left(J/\psi \rightarrow \text{V$\pi $}^0\right)} \times\\
&\sqrt{\frac{q^3\left(m_{\psi },\sqrt{s},m_{\pi ^0}\right)e^{-q^2\left.\left(m_{\psi },\sqrt{s},m_{\pi ^0}\right)\right/8\beta ^2}}{q^3\left(m_{\psi },m_V,m_{\pi ^0}\right)e^{-q^2\left.\left(m_{\psi },m_V,m_{\pi ^0}\right)\right/8\beta ^2}}}.
\end{split}
\end{equation}
\end{small}
%From Eq.~(\ref{eq:mysigma}) we obtain:
%\begin{equation}\label{eq:sigmapi0}
%\sigma_{\pi^0} =\sum _{i=1}^3 \text{par}_i^2\cdot \sigma _i+\sum _{i=1}^2 \sum _{j=i+1}^3 \text{par}_i\cdot \text{par}_j\cdot \sigma _{\text{ij}}
%\end{equation}

The values of $Br(\jpsi\to\omega\eta)$ and $Br(\jpsi\to\omega\eta^{\prime})$ can be calculated similarly as in $Br(\jpsi\to\omega\pi^0)$ case. They have the same form as in Eqs.~(\ref{eq:bruncor}), but Eq.~(\ref{eq:ajpsivp3}) has a little difference:
\begin{small}
\begin{equation}
\begin{split}
A_{\psi V\eta^{(\prime )}}(s)\equiv &\sqrt{Br\left(J/\psi \rightarrow \text{V$\eta $}^{(\prime )}\right)} \times \\
&\sqrt{\frac{q^3\left(m_{\psi },\sqrt{s},m_{\eta ^{(\prime )}}\right)e^{-q^2\left.\left(m_{\psi },\sqrt{s},m_{\eta ^{(\prime )}}\right)\right/8\beta ^2}}{q^3\left(m_{\psi },m_V,m_{\eta ^{(\prime )}}\right)e^{-q^2\left.\left(m_{\psi },m_V,m_{\eta ^{(\prime )}}\right)\right/8\beta ^2}}}.
\end{split}
\end{equation}
\end{small}

The branching ratios reported in PDG2012~\cite{PDG2012} are listed in the third column of Table ~\ref{tab:brvaluesfit}, subscript ``cor'' and ``uncor'' mean without and with the contribution of mixing effect, respectively.

In general 12 parameters appear in Table~\ref{tab:amplitudes1} and Table~\ref{tab:brvaluesfit}, they are $g$, $e$, $r$, $s$, $s_V$, $s_P$, $\theta$, $\phi_P$, $\phi_{\eta^{\prime} G}$, $r^{\prime}$, $f$ and $\Pi_{\rho\omega}$. However, we got 11 branching ratios in Table~\ref{tab:brvaluesfit}.
Fixing some parameters to the expected values~\cite{Escribano2010,Thomas2007}, we may fit the remaining parameters by minimizing
\begin{equation}
\chi ^2=\frac{1}{N}\sum _i \frac{\left(Br_i^{\text{vis}}-Br_i^{\text{th}}\right)^2}{\Delta _i^2},
\end{equation}
where $Br_i^{\text{vis}}$ and $\Delta _i$ are the $J/\psi\to VP$ branching ratios and corresponding errors given by PDG2012~\cite{PDG2012}; and $Br_i^{\text{th}}$ is calculated by Eq.~(\ref{eq:brcor}), except $Br(\jpsi\to\omega\pi^0(\eta, \eta^{\prime}))$ which is calculated by Eq.~(\ref{eq:bruncor}). $N$ is the number of branching ratios used.

The fit is performed according to following configuration which needs to be defined. We mark all items as ``tag'' and each item ``tag[i]'' is described below:
\begin{itemize}
\item{tag[1]: defines whether $\rho-\omega$ mixing is taken into account in the fit. If $\rho-\omega$ mixing is not included, we just need to fit with Table~\ref{tab:amplitudes1} and Eq.~(\ref{eq:brcor}), which is similar as in Refs.~\cite{Escribano2010}. ``tag[1] ''=1 or 2 refers to without or with mixing in fit,  respectively.}
\item{tag[2]: defines the initial values and step-width. ``tag[2] ''=1 or 2 refers to using reference values~\cite{Escribano2010,Thomas2007} as initial values and $0.01\%$ of them as step-widths, or set to ``0'' or ``1'' as initial values and $10^{-6}$ as step-widths respectively.}
\item{tag[3]: defines weather limit the parameters in a physis range. ``tag[3] ''=1 or 2 refer to no limit or limit respectively.}
\item{tag[4]: defines how to deal with the effects of the SU(3)-breaking contributions $s_V$ and $s_P$. ``tag[4] ''=1, 2 or 3 means  free in fit, fix to 0, or set to reference values~\cite{Escribano2010}. }
\item{tag[5]: defines how to deal with the contribution of gluonium $\phi_{\eta^{\prime} G}$ and $r^{\prime}$. ``tag[5] ''=1, 2 or 3 means  free in fit, fix to 0, or set to reference values~\cite{Escribano2010}.}
\item{tag[6]: defines whether the values of parameters $g$, $e$, $r$, $s$, $s_V$, $s_P$, $\theta$, $\phi_P$, $\phi_{\eta^{\prime} G}$ and $r^{\prime}$ are fixed to the values in Refs.~\cite{Escribano2010,Thomas2007}, then fit $f$ and $\Pi_{\rho\omega}$. ``tag[6] ''=1 or 2 refer to do not fix or fix those parameters respectively.}
\end{itemize}

%\end{multicols}
\input{./tables/tabPDGfitresultfinal20130611.tex}
%\begin{multicols}{2} 
The fit configuration is represented by the setting of these tag numbers. For example, ``tag=121211'' means no $\rho-\omega$ mixing; set ``0'' or ``1'' as initial values and $10^{-6}$ as step-width; no limits on parameters; $s_V=0, s_P=0$; $\phi_{\eta^{\prime} G}$ and $r^{\prime}$ are free; parameters are not fixed in the fit.

\subsection{Result of the Fit}
Two models have been used in fit: with form factor of $\jpsi$ (i.e. $\beta=0.5 \text{GeV}$) or without form factor of $\jpsi$ (i.e. $\beta=10^{10} \text{GeV}$). If a fit result does not satisfy $g>0$, $e>0$, $|r|<1$, $|s|<1$, $|s_V|<1$, $|s_P|<1$, $|r^{\prime}|<1$ and $\Pi_{\rho\omega}>0$, it has no physics meaning and is marked as ``Invalid''. The fit with $\chi^2/d.o.f<1.5$ is acknowledged as good fit. The results of good fits with valid physics meaning are studied carefully.

A detailed analysis described in next section shows that, it is much more resonable to take into account of  $\rho-\omega$ mixing and $\jpsi$ form factor (i.e. $\beta=0.5 \text{GeV}$) in the fit , the corresponding fit results are listed in Table~\ref{tab:PDGfitresultfinal}.

\subsection{Discussion}
We have following observations from the fit results of good fits with valid physics meaning:
\begin{itemize}
\item{Regardless of considering the form factor of $\jpsi$ or not, about half of 77 fit configurations give result with a reasonable $\chi^2$ value ($\chi^2/d.o.f<1.5$); }
\item{regardless of including mixing or not, most of fit results are consist with the results in Refs.~\cite{Thomas2007,Escribano2010}; }
\item{the fitted SU(3)-breaking contributions is very small with significant error, that is $s_V=0.03\pm 0.12,\ s_P=0.17\pm 0.17$; }
\item{the gluonium contribution has little effect on the fit. If it is considered, the fit results consist with Ref.~\cite{Escribano2010} $\phi_{\eta^{\prime} G}=32\pm 11,\ r^{\prime}=-0.04\pm 0.20$, especially when $\rho-\omega$ mixing is included; }
\item{the fit doesn't depend on whether a physics range limit is applied on parameters; }
\item{in case of ignoring $\rho-\omega$ mixing, there is no difference between setting fit initial values to be references values~\cite{Escribano2010,Thomas2007}, or generally used values (``0'' or ``1''); }
\end{itemize}

From the comparison between the cases $\beta=0.5 \text{GeV}$ and $\beta=10^{10} \text{GeV}$ we note that: taking  $\rho-\omega$ mixing into account, the fit can succeed in both cases of setting fit initial values to Ref.~\cite{Escribano2010,Thomas2007} values and generally used values (``0 or 1'') when $\jpsi$ form factor (i.e. $\beta=0.5 \text{GeV}$) is considered, otherwise (i.e. $\beta=10^{10} \text{GeV}$) the initial values have to be set to references values~\cite{Escribano2010,Thomas2007} to ensure a good fit.

If $\beta=0.5 \text{GeV}$, it should also be pointed out that the $\chi^2$ of the fit is better when $\rho-\omega$ mixing is considered than not, although the obtained parameters may differ a little from Ref.~\cite{Escribano2010}. While if $\beta=10^{10} \text{GeV}$, we see that, the $\chi^2$ of the fits are worse when $\rho-\omega$ mixing is included, although the obtained parameters are similar as in Ref.~\cite{Escribano2010}.

In summary, about half of fit configurations give stable, consistent and reasonable ($\chi^2/d.o.f<1.5$) fit results. The effects of the SU(3)-breaking contributions is small ($s_V=0.031\pm 0.12,\ s_P=0.17\pm 0.17$). The contribution of gluonium has negligible effect on the fit, when included the results are consist with Ref.~\cite{Escribano2010} ($\phi_{\eta^{\prime} G}=32\pm 11,\ r^{\prime}=-0.04\pm 0.20$). It is preferable to include $\rho-\omega$ mixing and $\jpsi$ form factor effects, which leads to a reasonable and stable result. The fit configurations of ``tag=211231'' and ``tag=212121'' (the second and fifth row in Table~\ref{tab:PDGfitresultfinal}) are accepted, the branching ratios calculated according to the two sets of fitted parameters are listed in the fifth (``Fit 2'') and fourth (``Fit 1'') column in Table~\ref{tab:brvaluesfit} respectively, their errors are evaluated by assuming the fitted parameters following Gaussian distribution, and randomly picking 1000000 points to calculate deviation to the observed branching ratios. Taking errors into account, the ``tag=211231'' configuration is preferred.

%% file: tables/amplitudes1_20130611.tex
\begin{table*}[!htbp]
\centering
\small
\caption{\label{tab:amplitudes1}General parametrization of amplitudes for $\jpsi\to VP$.}
%\begin{ruledtabular}
\begin{tabular}{llll}
\hline \hline
\textrm{Process} &
\textrm{Amplitude ($M_{i})$} \\
\hline
$\rho ^+\pi ^-, \rho ^0\pi ^0, \rho ^-\pi ^+$ &
$g+e E^{i\theta}$  \\

$K^{*+}K^-, K^{*-}K^+$ &
$g(1-s)+e E^{i\theta}\left(1+s_e\right)$  \\

$K^{*0}\bar{K}^0, \bar{K}^{*0}K^0$ &
$g(1-s)-e E^{i\theta}\left(2-s_e\right)$  \\

$\omega_q \eta$ &
$\left(g+e E^{i\theta}\right) X_{\eta }+\sqrt{2}r g \left[\sqrt{2}X_{\eta }+\left(1-s_P\right)Y_{\eta }\right]+\sqrt{2}r' g Z_{\eta }$ \\

$\omega_q \eta '$ &
$\left(g+e E^{i\theta}\right) X_{\eta '}+\sqrt{2}r g \left[\sqrt{2}X_{\eta '}+\left(1-s_P\right)Y_{\eta '}\right]+\sqrt{2}r' g Z_{\eta '}$ \\

$\phi_s \eta$ &
$\left[g(1-2s)-2e E^{i\theta}\left(1-s_e\right)\right] Y_{\eta }+r g\left(1-s_V\right)\left[\sqrt{2}X_{\eta }+\left(1-s_P\right)Y_{\eta }\right]+r' g\left(1-s_V\right)Z_{\eta }$ \\

$\phi_s \eta '$ &
$\left[g(1-2s)-2e E^{i\theta}\left(1-s_e\right)\right] Y_{\eta '}+r g\left(1-s_V\right)\left[\sqrt{2}X_{\eta '}+\left(1-s_P\right)Y_{\eta '}\right]+r' g\left(1-s_V\right)Z_{\eta '}$ \\

$\rho \eta$ &
$3eE^{i\theta}X_{\eta }$ \\

$\rho \eta '$ &
$3eE^{i\theta}X_{\eta '}$ \\

$\omega_q \pi ^0$ &
$3e E^{i\theta}$ \\

$\phi_s \pi ^0$ &
$0$ \\
\hline \hline
\end{tabular}
%\end{ruledtabular}
\end{table*}

%% file: tables/brvaluesfit_20130611.tex
\begin{table*}[!htbp]
\centering
\small
\caption{\label{tab:brvaluesfit}The branching ratios $\jpsi\to VP$ ($\times 10^{-3}$) from PDG2012 and from the fit. ``Fit 1'' and ``Fit 2'' are for two different fit parameter configurations described in the text.}
%\begin{ruledtabular}
\begin{tabular}{lllll}
\hline \hline
\textrm{No.} &
\textrm{Process} &
\textrm{PDG2012~\cite{PDG2012}} &
\textrm{Fit 1 ($\chi^2/d.o.f.=0.022/1$)}&
\textrm{Fit 2 ($\chi^2/d.o.f.=1.61/3$)}\\
\hline
1& $\rho ^+\pi ^- +\rho ^0\pi ^0 +\rho ^-\pi ^+$ &        $16.9\pm 1.5$ &  16.9     $\pm$ 1.2      & 15.93     $\pm$ 0.82         \\
2& $K^{*+}K^- +K^{*-}K^+$ &        $5.12\pm 0.30$ &                        5.12     $\pm$ 0.21     & 5.25      $\pm$ 0.14         \\
3& $K^{*0}\bar{K}^0 +\bar{K}^{*0}K^0$ &        $4.39\pm 0.31$ &            4.39     $\pm$ 0.19     & 4.54      $\pm$ 0.25         \\
4& $(\omega \eta)_{cor}$ &        - &                                      1.279    $\pm$ 0.050    & 1.48      $\pm$ 0.17         \\
5& $(\omega \eta^{\prime})_{cor}$ &        - &                                0.13     $\pm$ 0.26     & 0.155     $\pm$ 0.056        \\
6& $\phi \eta$ &        $0.75\pm 0.08$ &                                   0.86     $\pm$ 0.13     & 0.79      $\pm$ 0.10         \\
7& $\phi \eta^{\prime}$ &        $0.40\pm 0.07$ &                             0.38     $\pm$ 0.21     & 0.370     $\pm$ 0.066        \\
8& $\rho \eta$ &        $0.193\pm 0.023$ &                                 0.1930   $\pm$ 0.0043   & 0.1968    $\pm$ 0.0040       \\
9& $\rho \eta^{\prime}$ &        $0.105\pm 0.018$ &                           0.105    $\pm$ 0.024    & 0.100     $\pm$ 0.018        \\
10& $(\omega \pi ^0)_{cor}$ &        - &                                   0.320    $\pm$ 0.032    & 0.364     $\pm$ 0.037        \\
11& $\phi \pi ^0$ &        $<6.4\times 10^{-3}(\text{C.L.}90\%)$ &                  0.00095  $\pm$ 0.00020  & 0.00108   $\pm$ 0.00021      \\
12& $(\omega \pi ^0)_{uncor}$ &        $0.45\pm 0.05$ &                    0.45     $\pm$ 0.93     & 0.45      $\pm$ 0.25         \\
13& $(\omega \eta)_{uncor}$ &        $1.74\pm 0.20$ &                      1.74     $\pm$ 0.45     & 1.72      $\pm$ 0.41         \\
14& $(\omega \eta^{\prime})_{uncor}$ &        $0.182\pm 0.021$ &              0.18     $\pm$ 0.18     & 0.184     $\pm$ 0.036        \\
\hline \hline
\end{tabular}
%\end{ruledtabular}
\end{table*}

%% file: tables/tabPDGfitresultfinal20130611.tex
\begin{table*}[!htbp]
\centering
\caption{\label{tab:PDGfitresultfinal}Result of fit with $\rho-\omega$ mixing and $\jpsi$ form factor effects, i.e. $\beta=0.5 \text{GeV}$ ($\chi^2/d.o.f<1.5$).
The index of the fit (in first column) marked with ``$*$'' means the fit results have large difference with the values in references (listed in the first row). ``Dif'' is defined as $\text{Dif}=\sum{(|x_{fit}-x_{Ref}|/\Delta(x)_{Ref})}$ , $x_{fit}$ and $x_{Ref}$  are the values of parameters ($g$, $e$, $r$, $s$, $s_V$, $s_P$, $\theta$, $\phi_P$, $\phi_{\eta^{\prime} G}$ and $r^{\prime}$) from the fit or from the reference respectively, $\Delta(x)_{Ref}$  is the error from the reference.
}
\tiny
%\begin{ruledtabular}
\begin{tabular}{lllllllllllllll}
\hline \hline
\textrm{No.} &\textrm{tag/Dif} &\textrm{$ g$ } &\textrm{$ e$ } &\textrm{$ r$ } &\textrm{$ s$ } &\textrm{$ s_V$ } &\textrm{$ s_P$ } &\textrm{$ s_e$ } &\textrm{$ \theta$ } &
\textrm{$ \phi_P$ } &\textrm{$ \phi_{\eta^{\prime} G}$ } &\textrm{$ r_P$ } &\textrm{$ \prod_{\rho\omega}(\text{ \text{GeV}}^2)$ } &\textrm{$ f(\text{ \text{GeV}}^2)$ }  \\

        &\textrm{$ \chi^2 / (d.o.f)$ }& \textrm{err}&\textrm{err}&\textrm{err}&\textrm{err}&\textrm{err}&\textrm{err}&\textrm{err}&
        \textrm{err}&\textrm{err}&\textrm{err}&\textrm{err}&\textrm{err}&\textrm{err}\\

\hline
&Ref~\cite{Thomas2007,Escribano2010}/0  &2.11    &0.213   &-0.43   &0.27    &-0.03   &-0.08   &0.19    &1.34    &44.6    &32      &-0.04   &  &     \\
&2.6 /3  &0.10    &0.012   &0.08    &0.03    &0.09    &0.10    &0.05    &0.12    &4.1     &11      &0.20    &            &     \\

\hline

1*& 211221/11.45 &2.200     &0.1800    &-0.350   &0.300     &0       &0       &0.19    &1.30     &38.0      &0       &0       &0.0140   &0.00410  \\
   &   1.7   /3      &0.073   &0.0090   &0.012   &0.022   &0       &0       &0       &0.13    &2.6     &0       &0       &0.0063  &0.00050  \\

\hline

2& 211231/5.375 &2.200     &0.2000     &-0.390   &0.290    &0       &0       &0.19    &1.30     &42.0      &32      &-0.04   &0.006  &0.00450  \\
   &   1.61  /3      &0.077   &0.0099  &0.014   &0.025   &0       &0       &0       &0.12    &2.7     &0       &0       &0.011   &0.00053 \\

\hline

3& 211321/9.779 &2.200     &0.1800    &-0.340   &0.290    &-0.03   &-0.08   &0.19    &1.30     &38.0      &0       &0       &0.0170   &0.00380  \\
   &   3.02  /3      &0.073   &0.0090   &0.012   &0.023   &0       &0       &0       &0.13    &2.6     &0       &0       &0.0059  &0.00047 \\

\hline

4& 211331/3.742 &2.200     &0.190    &-0.380   &0.280    &-0.03   &-0.08   &0.19    &1.30     &41.0      &32      &-0.04   &0.0110   &0.00410  \\
   &   3.04  /3      &0.078   &0.010    &0.014   &0.025   &0       &0       &0       &0.12    &2.8     &0       &0       &0.0083  &0.00049 \\

\hline

5*& 212121/14.55 &2.20     &0.1800    &-0.360   &0.320    &0.03   &0.17    &0.19    &1.30     &38.0      &0       &0       &0.004  &0.0052  \\
   &   0.022 /1      &0.10     &0.0090   &0.029   &0.036   &0.12    &0.17    &0       &0.13    &2.9     &0       &0       &0.023   &0.0014  \\

\hline

6& 212131/7.811 &2.200     &0.1900    &-0.400    &0.310    &0.022   &0.11    &0.19    &1.30     &42.0      &32      &-0.04   &0.000       &0.00500   \\
   &   0.454 /1      &0.099   &0.0091  &0.020    &0.032   &0.089   &0.10     &0       &0.12    &2.8     &0       &0       &0.041   &0.00062 \\

\hline

7& 212231/5.375 &2.200     &0.200     &-0.390   &0.290    &0       &0       &0.19    &1.30     &42.0      &32      &-0.04   &0.006  &0.00450  \\
   &   1.61  /3      &0.077   &0.010    &0.013   &0.024   &0       &0       &0       &0.12    &2.7     &0       &0       &0.011   &0.00054 \\

\hline

8& 212321/9.779 &2.200     &0.1800    &-0.340   &0.290    &-0.03   &-0.08   &0.19    &1.30     &38.0      &0       &0       &0.0170   &0.00380  \\
   &   3.02  /3      &0.073   &0.0090   &0.012   &0.023   &0       &0       &0       &0.13    &2.6     &0       &0       &0.0058  &0.00046 \\

\hline

9& 212331/3.742 &2.200     &0.190    &-0.380   &0.280    &-0.03   &-0.08   &0.19    &1.30     &41.0      &32      &-0.04   &0.0110   &0.00410  \\
   &   3.04  /3      &0.078   &0.010    &0.014   &0.025   &0       &0       &0       &0.12    &2.8     &0       &0       &0.0083  &0.00049 \\

\hline

10*& 221131/37.6  &2.20     &0.2000     &-0.690   &0.320    &0.500     &0.17    &0.19    &1.30     &-42.0     &32      &-0.04   &0.035   &0.0019  \\
   &   0.029 /1      &0.10     &0.0099  &0.062   &0.036   &0.056   &0.18    &0       &0.12    &2.9     &0       &0       &0.018   &0.0010   \\

\hline

11& 222331/3.742 &2.200     &0.1900    &-0.380   &0.280    &-0.03   &-0.08   &0.19    &1.30     &41.0      &32      &-0.04   &0.0110   &0.00410  \\
   &   3.04  /3      &0.077   &0.0099  &0.014   &0.025   &0       &0       &0       &0.12    &2.7     &0       &0       &0.0082  &0.00049 \\
\hline \hline
\end{tabular}
%\end{ruledtabular}
\end{table*}

%% file: charpters/Conclusion20130605.tex
\section{CONCLUSIONS}
From the global fit to PDG data according to our theoretical frame describing $\jpsi\to VP$ process, we obtained the parameters of the flavor parameterization model as listed in Table~\ref{tab:PDGfitresultfinal}. It turns out that whether the contribution of gluonium is considered has little effect on the fit, if considered the fit gives consistent results with the values in Ref.~\cite{Escribano2010} ($\phi_{\eta^{\prime} G}=32\pm 11,\  r^{\prime}=-0.04\pm 0.20$). The effects of the SU(3)-breaking contributions are also negligible:
\begin{equation}
s_V=0.03\pm 0.12, \ \
s_P=0.17\pm 0.17.
\end{equation}

Including mixing effect in the fit, we renewed the branching ratios of $Br(J/\psi\to \omega\pi^0(\eta, \eta^{\prime}))$ as listed in Table~\ref{tab:brvaluesfit}, It should be noted that, about $19\% \ (15\%,\ 15\%)$ difference in branching ratios from PDG2012 values~\cite{PDG2012} are observed when mixing effects are incorporated.
\begin{equation}
\begin{split}
&Br(J/\psi\to \omega\pi^0) = (3.64 \pm 0.37)\times 10^{-4},\\
&Br(J/\psi\to \omega\eta)  = (1.48 \pm 0.17)\times 10^{-3},\\
&Br(J/\psi\to \omega\eta^{\prime}) = (1.55\pm 0.56)\times 10^{-4}.
\end{split}
\end{equation}

The value of $\rho-\omega$ mixing polarization operator is also obtained:
\begin{equation}
\Pi _{\rho \omega }=0.006\pm 0.011\text{ \text{GeV}}^2,
\end{equation}
 The significance of $\Pi _{\rho \omega }$ is $0.36$, which means that it has a large probability to be zero. This value is comparable with the value calculated by formula~\cite{Achasov2003}:
\begin{equation}
\begin{split}
\Pi _{\rho \omega }&=\text{Re}\left(\Pi _{\rho \omega }\right)=\sqrt{\frac{\Gamma _{\omega }}{\Gamma _{\rho ^0}\left(m_{\omega }\right)}Br\left(\omega \rightarrow \pi ^+\pi ^-\right)}\times \\
                &\quad \times \left|\left(m_{\omega }^2-m_{\rho ^0}^2\right)-i m_{\omega }\left(\Gamma _{\omega }-\Gamma _{\rho ^0}\left(m_{\omega }\right)\right)\right|.
\end{split}
\end{equation}
$\Pi _{\rho \omega }$ value will be 0.0042  $\text{GeV}^2$ or 0.0033  $\text{GeV}^2$ when parameters from SND~\cite{Achasov2003} or PDG2012~\cite{PDG2012} are used, respectively.\\

\begin{figure}
\includegraphics[scale=1.0,height=0.2\textheight,width=0.4\textwidth]
{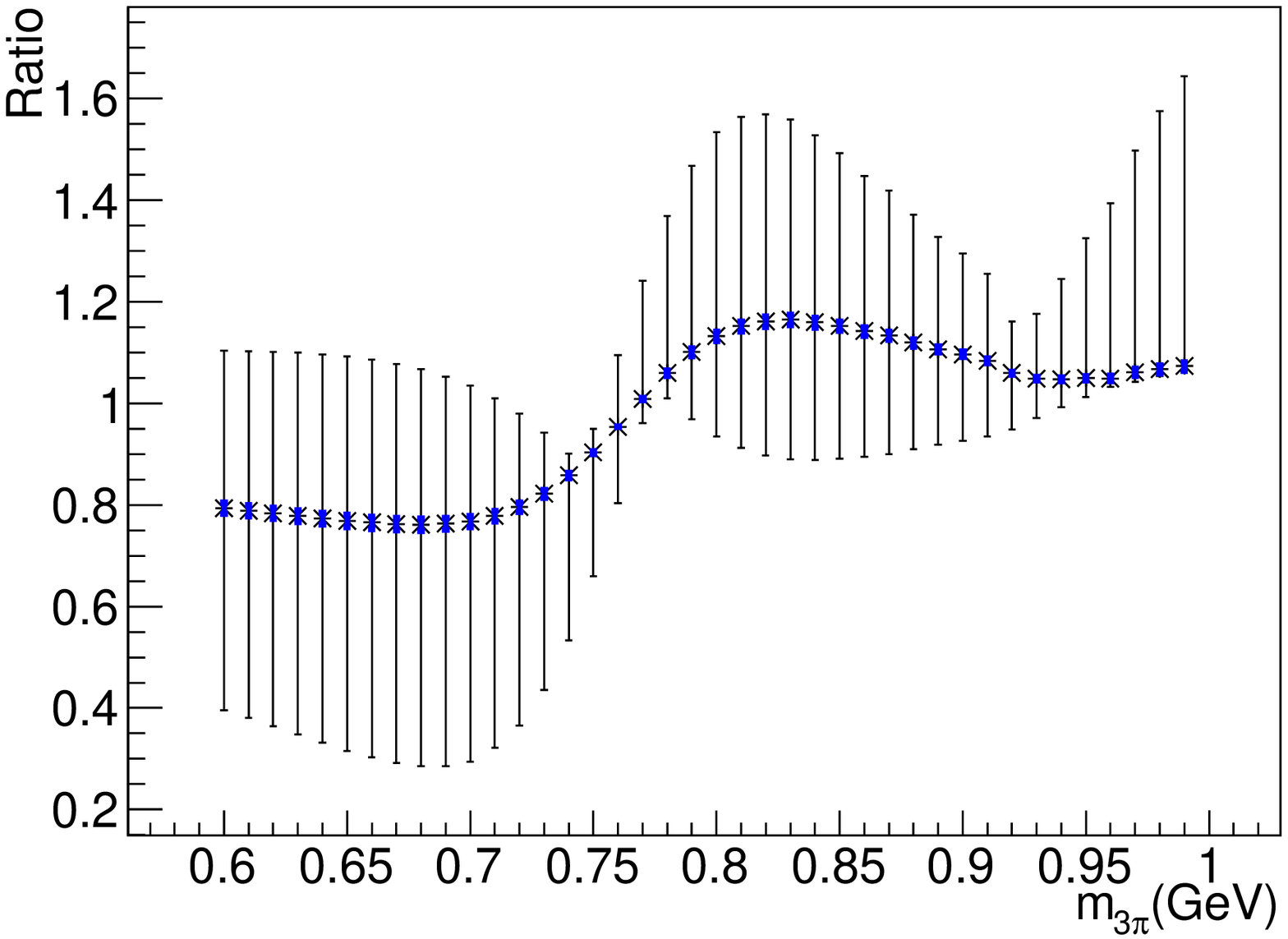}
\caption{\label{fig:ratioofsigam0}\small{The ratio of $\sigma(s)_{mix}$ with $\rho-\omega$ mixing ($\Pi _{\rho \omega }=0.006\text{\text{GeV}}^2$) to $\sigma(s)_{nomix}$ without $\rho-\omega$ mixing ($\Pi _{\rho \omega }=0\text{\text{GeV}}^2$).}}
%\end{center}
%\begin{center}
\includegraphics[scale=1.0,height=0.2\textheight,width=0.4\textwidth]
{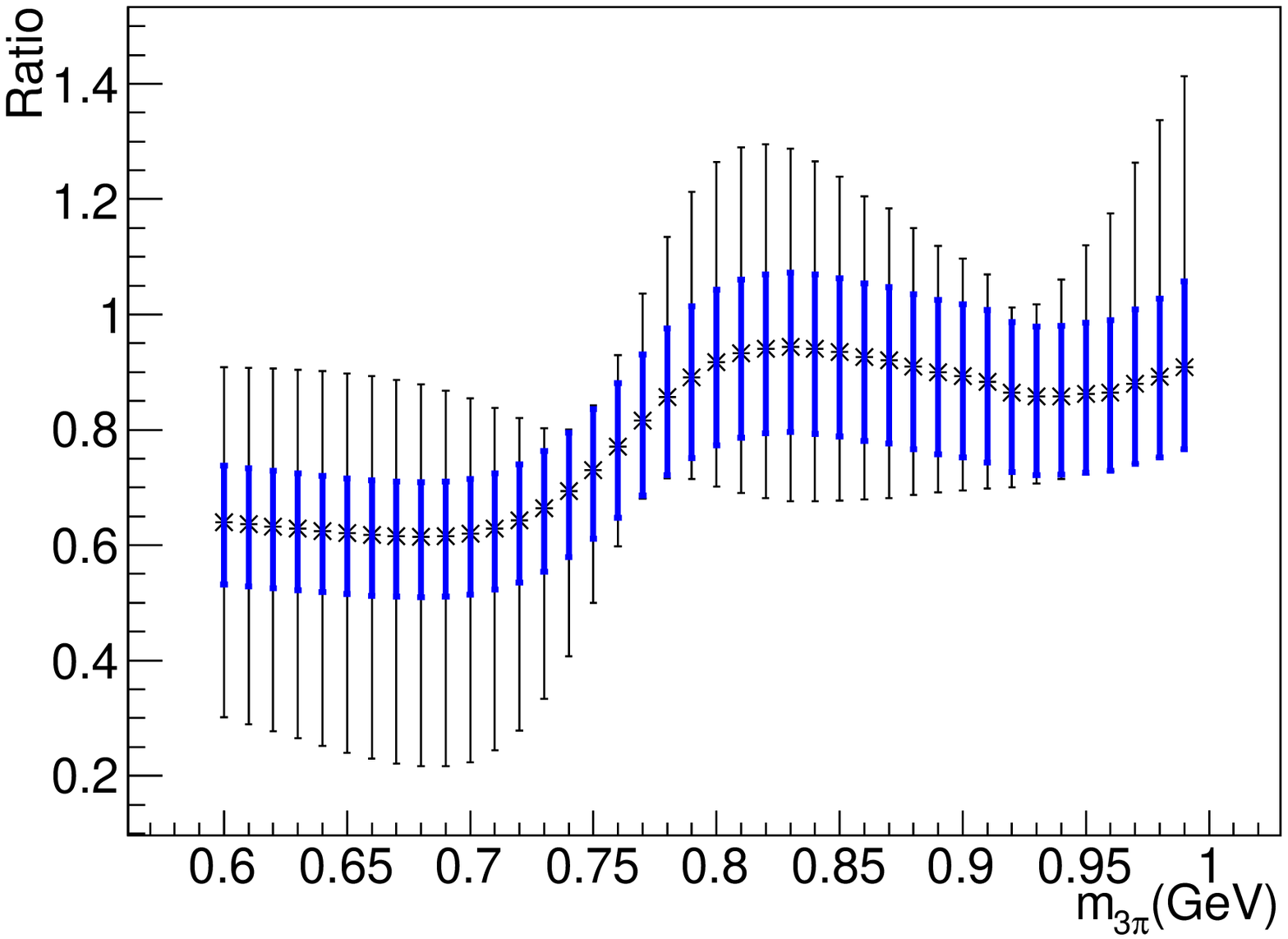}
\caption{\label{fig:ratioofsigam1}\small{The ratio of $\sigma(s)_{new}$ with our corrected branching ratios and mixing to $\sigma(s)_{old}$ with PDG2002's branching ratios and no mixing. Thick blue lines (color online) represent the errors calculated by ignoring error of $\Pi _{\rho \omega }$.}}
\end{figure}

Fig.~\ref{fig:ratioofsigam0} shows the ratio between the cross sections with ($\Pi _{\rho \omega }=0.006\text{\text{GeV}}^2$) and without ($\Pi _{\rho \omega }=0\text{\text{GeV}}^2$) $\rho-\omega$ mixing ($\sigma(s)_{mix}/\sigma(s)_{nomix}$) as a function of invariant mass of $3\pi$ system, where the corrected branching ratios are used. It can be seen clearly that the $\rho-\omega$ mixing has significant effect on the shape of $m_{3\pi}$ spectrum, the variance reaches about $\pm 20\%$ at the largest above or below $\omega$ nominal mass.

Fig.~\ref{fig:ratioofsigam1} shows the ratio between the cross sections with our corrected branching ratios (and with mixing) and with PDG2002's branching ratios (and no mixing) ($\sigma(s)_{new}/\sigma(s)_{old}$) as a function of invariant mass of $3\pi$ system. It can also be observed clearly that our derivation has significant effect on the shape of $m_{3\pi}$ spectrum, the variance reaches about $40\%$ at the largest and about $20\%$ nearby $\omega$ nominal mass.

The errors in Figs.~\ref{fig:ratioofsigam0} and \ref{fig:ratioofsigam1}
are caused mainly by the uncertainty of $\Pi _{\rho \omega }$, which has a limited significance. Further check is expected by the experiment.

%% file: charpters/Appendixes20130605.tex
\renewcommand{\theequation}{A\arabic{equation}}
\subsection*{Appendices A} \label{appendixesNotation}
\begin{small}
\noindent{\bf Notation in $\rho-\omega$ mixing}

The mechanism of $\rho-\omega$ mixing was reviewed in many references~\cite{OConnell1997,Yan2002,Reay1973,Achasov2005,Achasov2003,Achasov2002a,Achasov2002,Dimova2008}. The wave-functions of unmixed $\omega$ and $\rho$ states are given as~\cite{Achasov1992}:
\begin{equation}
\begin{split}
&|\omega^{(0)}\rangle \equiv\cfrac{1}{\sqrt{2}}|u\overline{u}+d\overline{d}\rangle, \\
&|\rho^{(0)}\rangle \equiv\cfrac{1}{\sqrt{2}}|u\overline{u}-d\overline{d}\rangle,
\end{split}
\end{equation}
while the wave-functions of physical states $\omega$ and $\rho$ under the pole approximation assumption can be written in general as:
\begin{equation}\label{eq:mixing1}
\begin{split}
&|\omega\rangle=|\omega^{(0)}\rangle+\varepsilon|\rho^{(0)}\rangle, \\
&|\rho\rangle=|\rho^{(0)}\rangle-\varepsilon|\omega^{(0)}\rangle,
\end{split}
\end{equation}
where the superscript $(0)$ denotes the coupling constants of the pure, unmixed states. Here
\begin{equation}\label{eq:mixing2}
\begin{split}
&\varepsilon=\cfrac{\prod_{\rho\omega}}{D_{\omega}(s)-D_{\rho}(s)},\\
&D_{V}(s)=m^{2}_{V}-s-i\sqrt{s}\Gamma_{V}(s).
\end{split}
\end{equation}
%Note, in Ref.~\cite{Achasov2003}:
%\begin{equation} %\label{eq:mixing2}
%\begin{split}
%&\varepsilon=\cfrac{-\prod_{\rho\omega}}{D_{\omega}(s)-D_{\rho}(s)}\\
%&D_{V}(s)=m^{2}_{V}-s-i\sqrt{s}\Gamma_{V}(s)
%\end{split}
%\end{equation}

$D_{V}(s)$ is the propagator function; $\Gamma_{V}(s)$ is the width of the vector meason; and $\prod_{\rho\omega}\equiv\langle\rho^{(0)}|W|\omega^{(0)}\rangle$~\cite{OConnell1997,Yan2002} is the polarization operator of the mixing. Note that $\varepsilon$ is not a real number, hence the teansfer matrix from isospin basis to physical basis is not unitary. In Ref.~\cite{Achasov2003} $\varepsilon$ is negative with the same expression.

Under this framework, the coupling constants for $\omega (\rho) \to \pi^{+}\pi^{-} (\rho\pi, 3\pi)$ decays can be determined as follows:
\begin{equation}\label{eq:mixing3}
\begin{split}
&g_{\omega\pi\pi}=g^{(0)}_{\omega\pi\pi}+\varepsilon g^{(0)}_{\rho\pi\pi},\qquad  g_{\rho\pi\pi}=g^{(0)}_{\rho\pi\pi}-\varepsilon g^{(0)}_{\omega\pi\pi}, \\
&g_{\omega\rho\pi}=g^{(0)}_{\omega\rho\pi} + \varepsilon g^{(0)}_{\rho\rho\pi}, \qquad g_{\rho\rho\pi}=g^{(0)}_{\rho\rho\pi}-\varepsilon g^{(0)}_{\omega\rho\pi}, \\
&g_{\omega 3\pi}=g^{(0)}_{\omega  3\pi}+\varepsilon g^{(0)}_{\rho  3\pi},\qquad  g_{\rho  3\pi}=g^{(0)}_{\rho  3\pi}-\varepsilon g^{(0)}_{\omega  3\pi},
\end{split}
\end{equation}
and for $J/\psi\to (\rho^{0}, \omega)\pi^0$:
\begin{equation}\label{eq:mixing4}
\begin{split}
&A_{\psi \omega \pi }(s)= A_{\psi \omega \pi }^{(0)}(s)+\varepsilon (s) A_{\psi \rho \pi }^{(0)}(s), \\
&A_{\psi \rho \pi }(s)= A_{\psi \rho \pi }^{(0)}(s)-\varepsilon (s) A_{\psi \omega \pi }^{(0)}(s),
\end{split}
\end{equation}
where ``g'' and ``A'' are defined as Section 2.\\

\noindent{\bf Notation in $\eta-\eta'$ mixing}

The wave-functions of physical states $\eta$ and $\eta'$ can be written in general as~\cite{Rosner1983,Escribano2010,Thomas2007,Kopke1989}:
\begin{equation}
\begin{split}
&|\eta\rangle=X_{\eta}|\eta_{q}\rangle+Y_{\eta}|\eta_{s}\rangle+Z_{\eta}|G\rangle, \\
&|\eta'\rangle=X_{\eta'}|\eta_{q}\rangle+Y_{\eta'}|\eta_{s}\rangle+Z_{\eta'}|G\rangle,
\end{split}
\end{equation}
where
\begin{equation}
\begin{split}
&|\eta_{q}\rangle\equiv\cfrac{1}{\sqrt{2}}|u\overline{u}+d\overline{d}\rangle, \qquad |\eta_{s}\rangle\equiv|s\overline{s}\rangle, \\
&|G\rangle=|gluonium\rangle,
\end{split}
\end{equation}
and
\begin{equation}
X^{2}_{\eta(\eta')}+Y^{2}_{\eta(\eta')}+Z^{2}_{\eta(\eta')}=1.
\end{equation}

Assume no gluonium content in $\eta$, the mixing can be parameterized in terms of two angles~\cite{Escribano2010,Thomas2007}:

\begin{equation} \label{eq:etaetapmixing}
\begin{split}
&X_{\eta}=\cos\phi_{P}, \qquad X_{\eta'}=\sin\phi_{P}\cos\phi_{\eta'G}, \\
&Y_{\eta}=-\sin\phi_{P}, \qquad Y_{\eta'}=\cos\phi_{P}\cos\phi_{\eta'G}, \\
&Z_{\eta}=0, \qquad \qquad  Z_{\eta'}=-\sin\phi_{\eta'G},
\end{split}
\end{equation}
where, $\phi_{P}$ is the $\eta-\eta'$ mixing angle, and $\phi_{\eta'G}$ weights the amount of gluonium in $\eta'$.\\

\noindent{\bf Notation in $\omega-\phi$ mixing}

Similar as in $\eta - \eta'$ mixing, a relatively simple expression for $\omega-\phi$ mixing is used~\cite{Escribano2010,Thomas2007}:

\begin{equation} \label{eq:omegaphimixing}
\begin{split}
&|\omega \rangle= \cos\phi_{\omega\phi}|\omega_{q}\rangle-\sin\phi_{\omega\phi}|\phi_{s}\rangle, \\
&|\phi \rangle= \sin\phi_{\omega\phi}|\omega_{q}\rangle+\cos\phi_{\omega\phi}|\phi_{s}\rangle,
\end{split}
\end{equation}
where $|\omega_{q} \rangle$ and $|\phi_{s} \rangle$ are the analog non-strange and strange states of $|\eta_{q} \rangle$ and $|\eta_{s} \rangle$ respectively, $\phi_{\omega\phi}$ is the mixing angle between $\rho$ and $\omega$. \\

\end{small}

%\subsection*{Appendices B}
%\noindent{\bf Fit Results} \label{appendixesPDGfitresultAll}
%%\end{multicols}
%%\vspace{10mm}
%%\input{./tables/tabPDGfitresultAll20130515.tex}
%%\vspace{10mm}
%%\input{./tables/tabPDGfitresultAllNoFF20130515.tex}
%\vspace{10mm}
%\input{./tables/tabPDGfitresult20130529.tex}
%\vspace{10mm}
%\input{./tables/tabPDGfitresultNoFF20130529.tex}
%%\begin{multicols}{2} 